\documentclass{icrc2009}

\usepackage{graphicx}   
\usepackage[caption=false]{caption}    
\usepackage[font=footnotesize]{subfig} 
\usepackage{fixltx2e}
\usepackage{url}
\usepackage{amssymb}
\usepackage{multirow}

\newcommand{\shorttitle}[1]%
{\markboth{Proceedings of the 31\MakeLowercase{$^{st}$} ICRC, {\L}\'{o}d\'{z} 2009}{#1} }
\newcommand{\etal}{\MakeLowercase{\textit{et al. }}} 

\begin{document}
\title{Experimental set-up of the LUNASKA lunar Cherenkov observations at the ATCA}
\author{\IEEEauthorblockN{Clancy James \IEEEauthorrefmark{1,2},
			  Ron Ekers \IEEEauthorrefmark{3},
                          Chris Philips\IEEEauthorrefmark{3},
                           Ray Protheroe\IEEEauthorrefmark{1},\\
                           Paul Roberts \IEEEauthorrefmark{3},
			   Rebecca Robinson \IEEEauthorrefmark{4},
			   Jaime Alvarez-Mu\~niz \IEEEauthorrefmark{5},
			   Justin Bray \IEEEauthorrefmark{1}}
                            \\
\IEEEauthorblockA{\IEEEauthorrefmark{1}School of Chemistry \& Physics, Univ.\ of Adelaide, SA 5005, Australia}
\IEEEauthorblockA{\IEEEauthorrefmark{2} Department of Astrophysics, Radboud University Nijmegen, Postbus 9010, 6500 GL Nijmegen,
The Netherlands}
\IEEEauthorblockA{\IEEEauthorrefmark{3}Australia Telescope National Facility, Epping, NSW 1710, Australia}
\IEEEauthorblockA{\IEEEauthorrefmark{4}School of Physics, Univ.\ of Melbourne, VIC 3010, Australia}
\IEEEauthorblockA{\IEEEauthorrefmark{5}Depto. de F\'\i sica de Part\'\i culas \& Instituto Galego de F\'\i sica de Altas Enerx\'\i as, \\ Univ.\ de Santiago de Compostela, 15782 Santiago de Compostela, Spain }
}
\shorttitle{C.W.~James \etal LUNASKA observations -- method.}
\maketitle

\begin{abstract}\author{\IEEEauthorblockN{Clancy James \IEEEauthorrefmark{1,2},
			  Ron Ekers \IEEEauthorrefmark{3},
                          Chris Phillips\IEEEauthorrefmark{3},
                           Ray Protheroe Scott\IEEEauthorrefmark{1},
                           Paul Roberts \IEEEauthorrefmark{3},
			   Rebecca Robinson \IEEEauthorrefmark{4},
			   Jaime Alvarez-Mu\~niz \IEEEauthorrefmark{5},
			   Justin Bray \IEEEauthorrefmark{1}}
                            \\
\IEEEauthorblockA{\IEEEauthorrefmark{1}School of Chemistry \& Physics, Univ.\ of Adelaide, SA 5005, Australia}
\IEEEauthorblockA{\IEEEauthorrefmark{2} Department of Astrophysics, Radboud University Nijmegen, Postbus 9010, 6500 GL Nijmegen,
The Netherlands}
\IEEEauthorblockA{\IEEEauthorrefmark{3}Australia Telescope National Facility, Epping, NSW 1710, Australia}
\IEEEauthorblockA{\IEEEauthorrefmark{4}School of Physics, Univ.\ of Melbourne, VIC 3010, Australia}
\IEEEauthorblockA{\IEEEauthorrefmark{5}Depto. de F\'\i sica de Part\'\i culas \& Instituto Galego de F\'\i sica de Altas Enerx\'\i as, Univ.\ de Santiago de Compostela, 15782 Santiago de Compostela, Spain }
}
This contribution describes the experimental set-up implemented by the LUNASKA project at the Australia Telescope Compact Array (ATCA) to enable the radio-telescope to be used to search for pulses of coherent Cherenkov radiation from UHE particle interactions in the Moon with an unprecedented bandwidth, and hence sensitivity. Our specialised hardware included analogue de-dispersion filters to coherently correct for the dispersion expected of a ~nanosecond pulse in the Earth's ionosphere over our wide (600 MHz) bandwidth, and FPGA-based digitising boards running at 2.048~GHz for pulse detection. The trigger algorithm is described, as are the methods used discriminate between terrestrial RFI and true lunar pulses. We also outline the next stage of hardware development expected to be used in our 2010 observations.
 \end{abstract}

 
\section{Introduction}
\label{sec1}

The lunar Cherenkov technique, proposed by Dagkesamanskii and Zheleznykh~\cite{Dagkesamanskii}, is a method to detect ultra-high energy cosmic-rays (CR) and neutrinos ($\nu$). The technique uses radio-telescopes to observe Earth's Moon, and search for the coherent Cherenkov (radio) radiation produced -- via the Askaryan Effect \cite{Askaryan} -- from a high-energy particle cascade in the Moon's outer layers. This enables the entire visible surface of the Moon to be used as detector of particles at energies where the radiation (which arrives as a short-duration pulse) is sufficiently strong to be detectable. Although past experiments, first at Parkes \cite{Parkes}, and subsequently at Goldstone (GLUE) \cite{GLUE}, Kalyazin \cite{Kalyazin}, and Westerbork \cite{NuMoon}, have not recorded any confirmed detections, simulations indicate that a next-generation radio-telescope such as the SKA (Square Kilometre Array \cite{ska}) could probe the `guaranteed' flux of UHE $\nu$, and would likely be sensitive to the known flux of UHE CR \cite{JP}.

Effectively utilising a radio-telescope to search for $O\sim1$~ns duration pulses is extremely difficult however. The LUNASKA -- Lunar UHE Neutrino Astrophysics with the SKA -- project therefore aims to develop the technique to enable an instrument such as the SKA to be used at its full sensitivity for a UHE particle search. To this end, we have been using the Australia Telescope Compact Array (ATCA), a radio interferometer of six $22$~m antenna located in New South Wales, Australia, as a test-bed to develop experimental methods scalable to giant, broad-bandwidth radio arrays such as the SKA. In particular, we target the `high-frequency' ($\gtrsim 1$~GHz) regime, both because ATCA and the likely high-frequency component of the SKA are/will be arrays of small dishes, and because the NuMoon/LOFAR collaboration \cite{NuMoon} is conducting similar development for low-frequency observations. Here, we describe our experimental set-up and equipment, as used for two three-day observation periods in February and May 2008, which allowed as to use an unprecedented $600$~MHz bandwidth in the one data channel to search for these elusive particles (results of this search are presented in Ref.\ \cite{results}).

\section{Overview}
\label{sec2}

\begin{figure*}[!t]
 \centering
 \includegraphics[width=4in]{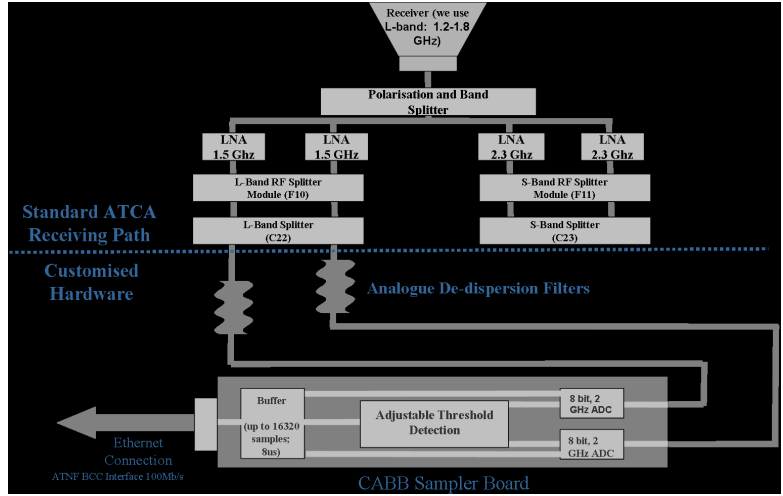}
 \caption{ATCA signal path, as implemented on antennas $1$, $3$, and $5$.}
 \label{fig01}
\end{figure*}

The ATCA is a suitable test-bed for the SKA in the few-GHz frequency range, where the SKA's collecting area will probably be a large number of small dishes. We used the ATCA's L-band receiver, with a nominal range of $1.2$--$1.8$ GHz, to give us $600$~MHz of bandwidth in dual linear `A' and `B' polarisations. At the time of the experiment, the cable to the control room could handle only $128$~MHz at our $8$-bit digitisation. Therefore, we used the sampler boards for the ATCA CABB (Compact Array BroadBand) upgrade to implement an FPGA-based digitisation and pulse detection algorithm at each of three of the six ATCA antennas. A diagram of the signal path at each antenna is given in Fig.\ \ref{fig01}.

The top half of Fig.\ \ref{fig01} describes the standard ATCA signal path. We used a maintenance plug to extract both polarised bands from this path, passed them through analogue dedispersion filters (see Sec.\ \ref{dedispersion}), and sent the signals to the CABB boards. These sampled each channel at $2.048$~GHz with $8$-bit precision, which was then copied both to a buffer of length up to $16,320$ samples ($\sim 8~\mu$s), and to the FPGA logic engine. Our trigger logic was necessarily simple, due to the difficulties of implementing complex logic in real time. A trigger occurred if the voltage magnitude on either polarisation exceeded an adjustable threshold, upon which both A and B buffers from that antenna were returned to the control room and recorded to disk, along with a time-stamp at sampling accuracy.

Since communication on short timescales between the antennas was impossible, each operated independently, which meant that the sensitivity of the experiment was limited by the threshold on each individual antenna. To compensate, we set the threshold on each antenna to be as low as possible. Since the instrumental dead-time was approximately $8$~ms per $1~\mu$s of buffer length, we reduced the length of the buffers returned to $256$~samples ($125$~ns) per polarisation band. Setting the threshold between $5.5$ and $6~\sigma$ kept the trigger rates to $40-50$~Hz per band, for a mean dead-time per antenna of $5$\%. Our $8$-bit precision was thus barely enough to allow small ($<0.1~\sigma$) adjustments of the threshold values while keeping a reasonable dynamic range between the trigger threshold and the ADC saturation point. We recommend $10$-bit precision if possible for future experiments.

\section{Sampling and Triggering Effects}
\label{sec3}

The ($100$\% linearly-polarised) Cherenkov radiation emitted by a high-energy particle cascade in the coherent frequency regime will be intrinsically similar to a band-limited impulse, of order a few nanoseconds' duration. To model the response of our equipment to the full range of pulse shapes -- which vary with the shower energy and the interaction geometry -- we use two reference pulses: a `completely coherent' pulse ($E(f) \propto f$) pulse, as from a $10^{20}$~eV cascade viewed at the Cherenkov angle; and a `maximally incoherent' pulse ($E(f) \propto f \exp{-C f^2}$), as from a $10^{23}$~eV cascade viewed far from the Cherenkov angle. The pulses were chosen such that the peak intrinsic electric field strengths would be identical. Detectable pulses with regolith-absorption terms ($\exp{-C_2 f}$) will fall in-between these two extremes.

Our simple detection algorithm was inefficient in two important ways. Firstly, by only using polarisation channels individually (as opposed to in combination, e.g.\ $X^2+Y^2$) to form a trigger, the full signal strength would be seen only if the signal polarisation happened to align with one of the channels. Since the signal paths within the antenna for the two polarisation bands were similar but not identical, extra cable would have had to be inserted to adjust for path differences to allow the more optimal real-time trigger algorithm to be used, which was not possible for the observations reported here. Secondly, our non-infinite sampling rate meant that we would inevitably miss the narrow peak of a real pulse. Although we would have sufficient information to reconstruct the true signal in the $1.2$--$1.8$~GHz range to arbitrary accuracy in post-processing (sampling greater than Nyquist), doing so was impossible in real time. Therefore our sensitivity to narrow pulses was reduced, although our over-sampling partially compensated for this effect.

\section{Ionospheric dispersion}
\label{dedispersion}

The free electrons in the Earth's ionosphere result in a frequency-dependent refractive index, which causes any broadband signal to be dispersed in time (path-bending is a lesser effect here). The degree of dispersion depends on the total electron content (TEC) along the line-of-sight (LOS), with the phase-delay $\Delta \phi$, and the time-delay $\Delta t$, at frequency $f$ given by Eqs.\ \ref{eq:phase_delay} and \ref{eq:time_delay} respectively:
\begin{eqnarray}
\Delta \phi~~{\rm (rad)} &=& 2.68 \pi TECU \frac{1~{\rm GHz}}{f} \label{eq:phase_delay}\\
\Delta t~~{\rm (ns)} &=& 1.34 TECU \left( \frac{1~{\rm GHz}}{f} \right)^2 \label{eq:time_delay}
\end{eqnarray}
where TECU are total electron content units ($10^{16}$ e$^{-}$/m$^2$). For typical values of the TEC ($O\sim10$~TECU), this resulted in a dispersion of $\sim 5$~ns across our $1.2--1.8$~GHz band. Since the ionosphere is dynamic, and the LOS TEC (or `slant' TEC, STEC) goes as $1/\sin \alpha$ ($\alpha$ the elevation angle), the ultimate goal will be to measure and digitally correct for the dispersion in real time (see R.~McFadden \etal for our proposed method). Since the computational requirements for McFadden's method were too great at the time of observation, our solution was to use analogue dedispersion filters \cite{paul_filters}. Consisting of a $\sim 1.2$~m tapered microwave waveguide, these were set to correct for a dispersion corresponding to $9.67$ TECU over our particular bandwidth. The group-delay and amplitude response is shown in Fig.\ \ref{fig03}. The small deviations from the trend are due to physical limitations on the size of the filter (finite length), and imperfections in the construction (finite commercial competency), and can be neglected for this experiment.

\begin{figure}[!t]
 \centering
 \includegraphics[width=2.5in]{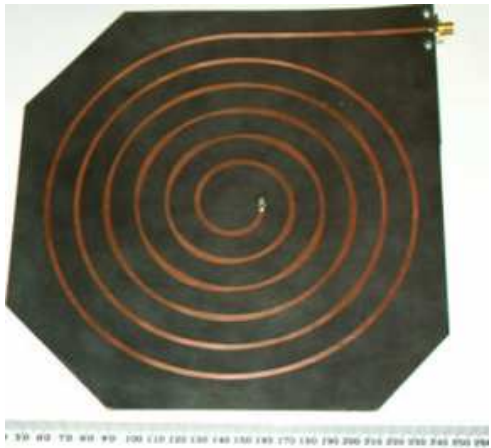}
 \caption{Photograph of our prototype dedispersion filter. The spiral shape is for compactness only -- the dedispersion is achieved by varying the waveguide width.}
 \label{fig02}
\end{figure}

\begin{figure}[!t]
 \centering
 \includegraphics[width=2.5in]{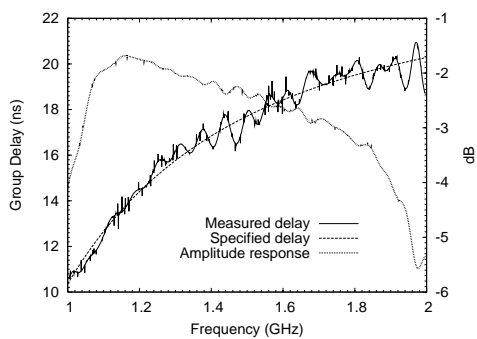}
 \caption{Phase delay and amplitude response of our analogue dedispersion filters.}
 \label{fig03}
\end{figure}

\begin{figure}[!t]
 \centering
 \includegraphics[width=2.5in]{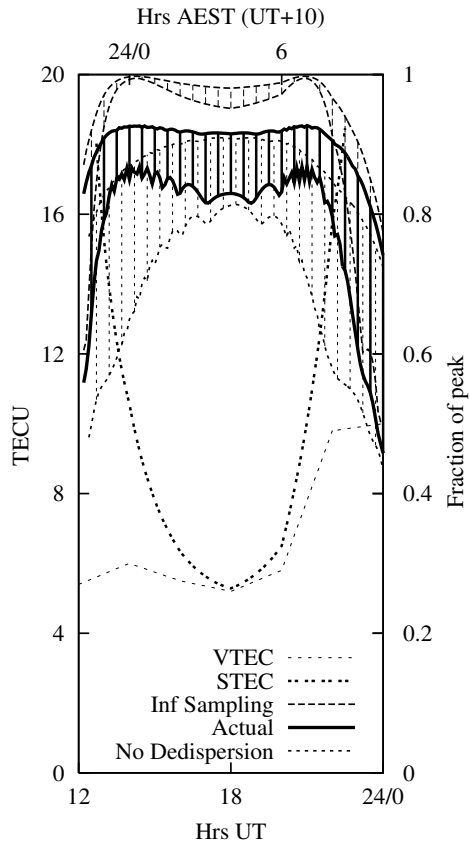}
 \caption{Fraction of peak signal -- loss is due to finite sampling and approximate dedispersion.}
 \label{fig04}
\end{figure}

Post-experiment, the vertical TEC (VTEC) was derived by interpolation of GPS measurements from Ref.\ \cite{NASA_crustal}. Including the lunar elevation angle gave the slant TEC, and the loss of sensitivity due to our approximate dedispersion and finite sampling rate calculated. The results for the (relatively low-VTEC) night of February $26^{\rm th}$ are plotted in Fig.\ \ref{fig04}. Also shown are the losses assuming an infinite sampling rate, and assuming no dedispersion filter. Our dedispersion method would have allowed on average $90$\% of the peak signal to be captured, with most of the remaining loss due to finite sampling. Without this method, only $80$\% of the peak signal would have been seen by the trigger algorithm.

\section{Time and Sensitivity Calibration}

Our unusual signal path meant that the automated methods of timing and sensitivity calibration between antennas would not work. For the timing calibration, we pointed the antennas predominantly at 3C273, set the buffers to their maximum length of $8~\mu$s, and triggered the buffers to within $1~\mu$s of each-other. For this, we used the noise diodes in the antennas, which emit pulses when they switch on. Correlation the common signal from 3C273 in the returned buffers then gave the time alignment.

For the sensitivitity calibration, we pointed the antennas at the Moon's centre, and set the trigger thresholds to zero, so as to capture an unbiased sample of received power. Repeating for an off-Moon pointing and subtracting the result left the lunar thermal emission as seen by the antennas. Modelling this as a $225$~K black body \cite{tt70}, and using measurements of the antenna beam pattern \cite{techmemo}, allowed us to extract the bandpass for each antenna.

\section{RFI}

Lunar Cherenkov experiments function as short-duration RFI monitoring experiments, since this is the `haystack' in which the `needles' of UHE particle signals will be found. Importantly, standard RFI monitoring experiments don't see the short-duration (and certainly not the sub-microsecond) RFI environment. Confusing such RFI for a lunar Cherenkov signal in offline processing is unlikely, since both fast timing between antennas, and the characteristic ionospheric dispersion effects, provide excellent discriminants. The danger lies however in these events saturating the trigger rate, so that the effective observation time drops to zero.

\begin{figure}[!t]
 \centering
 \includegraphics[width=2.5in]{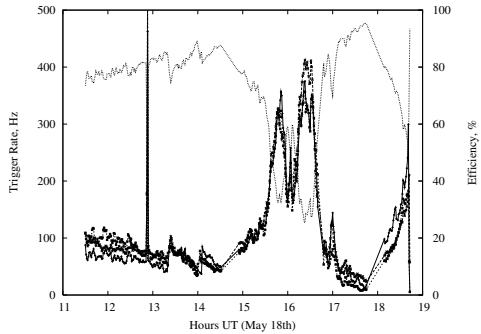}
 \caption{Trigger rates (lower lines) and efficiency (upper lines) from Eq.\ (\ref{ATCAtrig_rate_1}) for the night of May $18^{\rm th}$, 2008.}
 \label{figrates}
\end{figure}

We defined the efficiency of the experiment to be the fraction of observing time when all three antennas were sampling and ready to trigger, rather than `dead' and writing data from a previous trigger. For a trigger rate $r_i$ (Hz) on antenna $i$, saturation trigger rate $R_i$, and purely random trigger events, the efficiency $\xi$ is given by:
\begin{eqnarray}
\xi & = & \Pi_i \left(1-\frac{r_i}{R_i}\right) \label{ATCAtrig_rate_1}
\end{eqnarray}
where the $i$ multiplies over all three antennas. Our measured trigger rates and resulting efficiency are shown for one night of observations at the ATCA in Fig.\ \ref{figrates}. For most of the night, trigger rates (and hence efficiency) were well-behaved, with jumps being caused by adjustments in the thresholds and/or calibration periods. However, periods of RFI caused our rates to increase greatly, and for over an hour around $16:00$~UT, we ran at only $30$\% efficiency. Further analysis indicated that a significant fraction of this RFI was generated on-site, so that moving to a radio-quiet environment might be less effective than supposed. Another implication is that `incoherent' pulse searches, where outputs from multiple antennas are added into the one band to simplify the real-time logic, will be easily saturated by such RFI. Coherent addition of the signals in the voltage domain will be required to increase the power of lunar signals relative to the background if sensitivity is to be improved.

\vspace{2cm}

\section{Conclusion}

We have successfully implemented specialised pulse detection and dedispersion hardware at three of the six Australia Telescope Compact Array antennas. We estimate that the hardware would enable us to capture on average $86$\% of a pulse peak, which has enabled us to perform lunar Cherenkov pulse detection experiments using a very broad ($600$~MHz) continuous bandwidth, the results of which are given elsewhere \cite{results}. The limiting factor to our sensitivity was the difficulty of implementing real-time logic, and our inability to coherent add antenna voltages in the time domain. We expect to overcome these limitations in future observations using the full CABB system, and plans for using future radio instruments should be developed with these considerations in mind.

\section*{Acknowledgments}
The Australia Telescope Compact Array is part of the Australia Telescope which is funded by the Commonwealth of Australia for operation as a National Facility managed by CSIRO. This research was supported by the Australian Research Council's Discovery Project funding scheme (project
numbers DP0559991 and DP0881006). J.A-M thanks Xunta de Galicia (PGIDIT 06 PXIB 206184 PR) and
Conseller\'\i a de Educaci\'on (Grupos de Referencia Competitivos -- Consolider Xunta de Galicia 2006/51).

\end{document}